%% file: hbelusca_JJC2014_arxiv.tex
\documentclass[10pt,twoside,twocolumn]{book}

\usepackage[T1]{fontenc}
\usepackage[latin1]{inputenc}
\usepackage{geometry}
\usepackage{amssymb,amsmath,amsfonts}

\usepackage{nicefrac}
\usepackage{latexsym}
\usepackage{fancyhdr}
\usepackage{graphicx}
\usepackage{amsthm,multibox}
\usepackage{color}
\usepackage{hhline,xspace,rotate}
\usepackage{url}
\usepackage[french]{varioref}
\usepackage{colortbl}
\usepackage{verbatim}
\usepackage{caption}
\usepackage{subcaption}
\usepackage{braket}

\usepackage{import}
\usepackage{setspace}
\usepackage[breaklinks=true,debug=true]{hyperref}
\makeatletter

\newcommand \thearabicpart {\@arabic\c@part} 

\newlength{\picheight}
\newlength{\titlewidth}
\newsavebox{\picbox}
\savebox{\picbox}{%
  \parbox[b][3.6cm][c]{2.4cm}{}%
}

\def\institute#1{\gdef\@institute{#1}}
\def\@institute{\@latex@warning@no@line{No \noexpand\institute given}}
\def\maketitle{%
\refstepcounter{chapter}%
\twocolumn[
\null
\setlength{\titlewidth}{\textwidth}
\addtolength{\titlewidth}{-2\wd\picbox}
\addtolength{\titlewidth}{-3em}
\hskip\wd\picbox%
\hfill{\large\bf
  \parbox[b]
{\titlewidth}{%
    \centering\@title\\[2ex plus 2 fill]
    \Large\@author\\[0ex plus 1 fill]
    \large\it\@institute\\[2ex plus 2 fill]
    Proceedings of the 2014 Journées des Jeunes Chercheurs (JJC 2014), 7-13 December 2014, \\ Le Lazaret, Sète, France
  }%
}%
\hfill\usebox\picbox
\par%
\vskip 2em%
]%
\thispagestyle{plain}%
\addcontentsline{toc}{chapter}{{\let\\\protect\xspace\@author: \itshape\@title}}%
\setcounter{section}{0}%
}
\newenvironment{abstract}{\section*{Abstract}}{}

\def\@lbibitem[#1]#2{%
  \@skiphyperreftrue
  \H@item[%
    \ifx\Hy@raisedlink\@empty
      \hyper@anchorstart{cite.\thechapter.#2}\@BIBLABEL{#1}\hyper@anchorend
    \else
      \Hy@raisedlink{\hyper@anchorstart{cite.\thechapter.#2}\hyper@anchorend}%
      \@BIBLABEL{#1}%
    \fi
    \hfill
  ]%
  \@skiphyperreffalse
  \if@filesw
    \begingroup
      \let\protect\noexpand
      \immediate\write\@auxout{%
        \string\bibcite{\thechapter.#2}{#1}%
      }%
    \endgroup
  \fi
  \ignorespaces
}%
\def\@bibitem#1{%
  \@skiphyperreftrue\H@item\@skiphyperreffalse
  \Hy@raisedlink{\hyper@anchorstart{cite.\thechapter.#1}\relax\hyper@anchorend}%
  \if@filesw
    \begingroup
      \let\protect\noexpand
      \immediate\write\@auxout{%
        \string\bibcite{\thechapter.#1}{\the\value{\@listctr}}%
      }%
    \endgroup
  \fi
  \ignorespaces
}%
\def\@citex[#1]#2{%
  \let\@citea\@empty
  \@cite{\@for\@citeb:=#2\do
    {\@citea\def\@citea{,\penalty\@m\ }%
     \edef\@citeb{\thechapter.\expandafter\@firstofone\@citeb\@empty}%
     \if@filesw\immediate\write\@auxout{\string\citation{\@citeb}}\fi
     \@ifundefined{b@\@citeb}{\mbox{\reset@font\bfseries ?}%
       \G@refundefinedtrue
       \@latex@warning
         {Citation `\@citeb' on page \thepage \space undefined}}%
       {\hbox{\csname b@\@citeb\endcsname}}}}{#1}}

\renewenvironment{thebibliography}[1]
     {\section*{\refname}%
      \@mkboth{\MakeUppercase\refname}{\MakeUppercase\refname}%
      \list{\@biblabel{\@arabic\c@enumiv}}%
           {\settowidth\labelwidth{\@biblabel{#1}}%
            \leftmargin\labelwidth
            \advance\leftmargin\labelsep
            \@openbib@code
            \usecounter{enumiv}%
            \let\p@enumiv\@empty
            \renewcommand\theenumiv{\@arabic\c@enumiv}}%
      \sloppy
      \clubpenalty4000
      \@clubpenalty \clubpenalty
      \widowpenalty4000%
      \sfcode`\.\@m}
     {\def\@noitemerr
       {\@latex@warning{Empty `thebibliography' environment}}%
      \endlist}

\makeatother

\usepackage[frenchb, english]{babel}

\begin{document}

\def\rjcpartname{Beyond the Standard Model}
\pagestyle{fancy}
  {
    \cleardoublepage
    \subimport{session/hbelusca/}{hbelusca}  }

\end{document}

%% file: session/hbelusca/hbelusca.tex

\makeatletter

\newcommand*{\centernot}{%
  \mathpalette\@centernot
}
\def\@centernot#1#2{%
  \mathrel{%
    \rlap{%
      \settowidth\dimen@{$\m@th#1{#2}$}%
      \kern.5\dimen@
      \settowidth\dimen@{$\m@th#1=$}%
      \kern-.5\dimen@
      $\m@th#1\not$%
    }%
    {#2}%
  }%
}

\makeatother

\renewcommand{\imath}{i} 
\renewcommand{\Re}[1]{\operatorname{Re}(#1)} 
\renewcommand{\Im}[1]{\operatorname{Im}(#1)} 

\title{\textsc{Higgs Couplings in an Effective Theory Framework}}
\author{Hermès~\textsc{Bélusca--Maïto}}
\institute{Laboratoire de Physique Théorique, CNRS -- UMR 8627, \\ Bât. 210, Université Paris-Sud XI, F-91405 Orsay Cedex, France}

\maketitle

\begin{abstract}
	The study of the properties of the scalar boson recently discovered at the LHC~\cite{Aad:2012tfa,Chatrchyan:2012ufa} may allow us to know whether it is well described by the Standard Model. In the case where deviations from SM predictions are present, this would be an evidence for the presence of new physics. We focus on the study of the Higgs couplings to matter in a model-independent approach by introducing a dimension-6 effective Lagrangian that includes both CP-even and CP-odd effective couplings. Constraints are set on some of these coefficients using experimental data from ATLAS and CMS as well as electroweak precision measurements from LEP, SLC and Tevatron. These data meaningfully constrain CP-even and some CP-odd couplings.\footnote{Work done in collaboration with A. Falkowski (LPT Orsay).}
\end{abstract}

\section{Introduction}
	\par The Standard Model (SM) is a theory of matter based on the gauge group $SU(3)_C \times SU(2)_L \times U(1)_Y$. When writing the most general gauge-invariant Lagrangian with operators of mass-dimension up to 4 and compatible with experimental data, without doing any other hypotheses, no mass terms for the $W$ and $Z$ bosons and the fermions can be written because they would break gauge symmetry. It is however known experimentally that these particles have non-zero masses. Moreover this formulation of the SM suffers from the fact that longitudinal $W$--$W$ boson scattering amplitude is not unitary: it grows like the energy squared.

	\par Those issues can be solved by a mechanism found by R. Brout and F. Englert~\cite{Englert:1964et}, and independently by P. Higgs~\cite{Higgs:1964pj} and G. Guralnik, C.R. Hagen and T. Kibble~\cite{Guralnik:1964eu}. In this picture a complex scalar field $H$ doublet of $SU(2)_L$, dubbed the Higgs field, is introduced, and couples to the SM fermions via Yukawa interactions and to gauge bosons via the Higgs field kinetic term inducing also gauge-Higgs interactions:
	\begin{equation}
		\mathcal{L}_{SM} \supset + \left(Y_{ij} \overline{f^i_L} H f^j_R + \text{h.c.}\right) + \left| D_\mu H \right|^2 - V(H) \; .
	\end{equation}
	The Higgs field has the following quartic potential:
	\begin{equation}
		V(H) = \mu_H^2 H^\dagger H + \lambda (H^\dagger H)^2
	\end{equation}
	where $\mu_H^2$ is the Higgs mass parameter and $\lambda$ its self-coupling. When the neutral component of the Higgs field acquires a non-zero vacuum expectation value (vev), electroweak symmetry breaking (EWSB) happens. This can be realized if $\mu_H^2 < 0$ and $\lambda > 0$ (the positivity of $\lambda$ is required for stability of the potential). After EWSB the neutral component can be developed around its vev: $\frac{v}{\sqrt{2}} \left(1+\frac{h}{v}\right)$ where $h$ is the Higgs boson and $v$ is the Higgs vev: $v \approx 246~\text{GeV}$. This mechanism generates masses for the Higgs boson itself, for the EW gauge bosons and almost all the fermions, the neutrinos remaining massless.

	\par Despite having demonstrated strong success in explaining almost all experimental results, the SM is not completely satisfactory: it does not explain observed neutrino oscillations, it does not contain any viable dark matter candidate, it does not incorporate gravity nor explains the inflation of the Universe. In the Higgs sector, all the fermion masses are arbitrary as well as the mass of the Higgs boson itself. And the whole Higgs mechanism is ad-hoc. All those facts lead us to conclude that the SM is not the ultimate theory and thus we need to go beyond it.

	\par On the experimental side, the recent confirmation by the ATLAS and CMS experiments~\cite{Aad:2012tfa,Chatrchyan:2012ufa} of the existence of a scalar particle with mass $m_h \approx 126$~GeV and production cross-sections and decay rates compatible with those of the SM Higgs boson, triggered many studies of its properties. However, apart from this discovery, these experiments have not seen any direct evidence of new physics (NP) so far. It is expected to discover hints of NP at the next run of the LHC where the center-of-mass energy will increase from 8~TeV to 13 and 14~TeV, but for now, with the current LHC data, we are led to consider the following questions: \emph{What is the fundamental nature of the new discovered particle (and of EWSB)? What are its properties? Do its couplings to SM particles deviate from the predicted SM values?} For that, probing its properties with precision is mandatory. To answer these questions, there are mainly two approaches: either perform a study in the context of a specified model, or, use a model-independent approach via an effective theory framework. In the first approach we need to work inside a given NP model amongst many, with almost no experimental data to restrain the choice. On the contrary, since the SM describes very well experimental data and no NP particles have been yet found, this may signify that if they exist they should live at energy scales much higher than the EW scale. In that case, the effective theory framework is certainly more suited.

\section{Effective Theory Approach}
	\par The key point of the effective theory approach is that NP is assumed to appear at an energy scale $\Lambda_{NP}$ much higher than the energy scale currently probed by the experiments, which is around the EW scale. Its influence on physics well below the NP scale is to induce small deviations with respect to SM predictions. Formally speaking, for energies lower than $\Lambda_{NP}$ the NP fields are integrated-out and give rise to dimension higher than 4 non-renormalizable effective operators in the expansion of the effective Lagrangian:
	\begin{equation}
		\mathcal{L}_{eff} = \mathcal{L}_{SM} + \sum_{d \geq 5} \frac{\mathcal{C}^{(d)}}{\Lambda_{NP}^{d-4}} \mathcal{O}^{(d)}\left(\{\text{SM fields}\}\right)
	\end{equation}
	where the $\mathcal{C}^{(d)}$ are dimensionless effective couplings (Wilson coefficients) and the $\mathcal{O}^{(d)}$ are gauge-invariant local effective operators of mass-dimension $d \geq 5$ that are \emph{function of SM fields only}. The leading term in this expansion is the SM Lagrangian which contains operators up to dimension 4. At the level of dimension-5 operators there is only one respecting the SM gauge symmetry (Weinberg operator) which gives masses to neutrinos after EWSB and does not have any sizeable impact on Higgs phenomenology; however it violates lepton number conservation so it will not be considered in our study.

	\par A simple example of such an approach is the famous Fermi theory for $\beta$ or muon decay, which can be described in the low-energy SM (below the EW scale) with a 4-fermion interaction with the following vertex:
	\begin{equation}
		\frac{G_F}{\sqrt{2}} \overline{f_1} \gamma_\mu(1-\gamma_5) f_2 \times \overline{f_3} \gamma^\mu(1-\gamma_5) f_4 \; ,
	\end{equation}
	the $f_i$ being the involved fermions and $G_F$ the Fermi constant. In the full Standard Model theory, such an interaction is described at tree-level by the exchange of a virtual $W$ boson between the fermions, so that the amplitude writes:
	\begin{equation}
		\frac{g}{\sqrt{2}} \overline{f_1} \gamma^\mu\frac{1-\gamma_5}{2} f_2 \times \frac{g_{\mu\nu}-\frac{p_\mu p_\nu}{M_W^2}}{p^2-M_W^2} \times \frac{g}{\sqrt{2}} \overline{f_3} \gamma^\nu\frac{1-\gamma_5}{2} f_4 \; .
	\end{equation}
	What differs here is the presence of the $W$ boson propagator (written here in unitary gauge) and its couplings to the left-handed fermion currents with a strength proportional to the weak coupling $g$. In the SM the weak coupling $g$ has no mass-dimension as required. At low $p^2$ the $W$ propagator can be approximated\footnote{This is equivalent to "integrating-out" the $W$ boson by replacing it with the solution of its equation of motion in the SM Lagrangian.} by $\frac{1}{M_W^2}$ and gives rise to the effective coupling $\frac{g^2}{8 M_W^2} = \frac{G_F}{\sqrt{2}}$ which is nothing but the Fermi constant, of mass-dimension -2.

	\par For our matters we need to build an effective Lagrangian containing the SM Lagrangian plus operators of mass-dimension up to 6 that are made of SM fields only: a first requirement is that the Higgs boson $h$ is part of the Higgs field $H$ that transforms in the $(\textbf{1},\textbf{2},\frac{1}{2})$ representation of the SM gauge group $SU(3)_C \times SU(2)_L \times U(1)_Y$ and acquires an expectation value~$v$. Assuming baryon and lepton numbers conservation, a complete list of 59 operators was found in~\cite{Grzadkowski:2010es} regardless of fermion flavour. Expliciting these operators requires a choice of operator basis because they can be redefined into other ones using equations of motion. These operators can be classified into 7 categories\footnote{We use the covariant derivative $D_\mu = \partial_\mu - \imath g_V V_\mu$ that applies on the different fields, and the anti-Hermitian derivative $A^\dagger \overleftrightarrow{D_\mu} B \equiv A^\dagger (D_\mu B) - (D_\mu A)^\dagger B$.}: pure-gauge ($F_\mu^\nu F_\nu^\rho F_\rho^\sigma$) and 4-fermion ($\overline{f_1} f_2 \overline{f_3} f_4$) operators, 2-fermion-Higgs vertex ($[\overline{f_1} \gamma^\mu f_2] [H^\dagger \overleftrightarrow{D_\mu} H]$) and 2-fermion-Higgs dipole ($H \overline{f_1} \sigma^{\mu\nu} f_2 F_{\mu\nu}$) operators, gauge-Higgs ($|H^\dagger \overleftrightarrow{D_\mu} H|^2$ or $[H^\dagger H] F_{\mu\nu} F^{\mu\nu}$) operators, Yukawa-like ($[H^\dagger H] [\overline{f_1} H f_2]$) operators and pure-Higgs ($[\partial_\mu (H^\dagger H)]^2$ or $[H^\dagger H]^3$) operators. CP-odd counterparts of these operators are also present. For Higgs phenomenology purposes we keep only operators that contain at least one Higgs doublet and which can be currently constrained by LHC data.

\section{Towards the phenomenological Higgs Lagrangian}
\label{sec:PhenoHiggsL}
	\par In order to get meaningful constraints for Higgs couplings with the current LHC data, extra assumptions need to be imposed about the underlying physics. Amongst the operators that are involved in Higgs phenomenology, we ignore the $(H^\dagger H)^3$ term which only modifies the Higgs self-coupling, because current LHC precision is not enough to correctly constrain it. We assume the absence of any source of flavour violation: this means that for all the operators involving two fermions, the coupling matrix is taken to be diagonal in flavour space. Absence of 2-fermion vertex and dipole operators is also assumed: the reason is that they are not fully constrained by current LHC data but only indirectly via measurements of electric dipole moments, and they can be a source of flavour violation mediated by the Higgs boson. A study of such couplings is currently in progress~\cite{Belusca-Maito:InPreparation}. Finally only the gauge-Higgs operators and Yukawa-like operators with their CP-odd counterparts are kept, as well as the pure-Higgs $[\partial_\mu (H^\dagger H)]^2$ operator that gives corrections to the Higgs kinetic term\footnote{The full list of considered operators is given in~\cite{Belusca-Maito:2014dpa}.}.

	\par EW precision tests give strong constraints on all possible new physics that can affect radiative corrections to the EW gauge bosons propagators. They are customarily parametrized by the three Peskin-Takeuchi $S$, $T$, $U$ oblique parameters~\cite{Peskin:1991sw}. The effective Higgs Lagrangian introduces corrections to these parameters, and at loop-level logarithmic and quadratic divergences in $\Lambda_{NP}$ appear. However the oblique parameters are found to be small experimentally~\cite{Baak:2012kk}: at $U = 0$ one has: $S = 0.05 \pm 0.09$ and $T = 0.08 \pm 0.07$. Therefore the dominant divergences are required to cancel. Other considerations need to be taken into account~\cite{Belusca-Maito:2014dpa} to remove them completely. These requirements introduce non-trivial relationship between the effective couplings that can be interpreted as extended custodial relations~\cite{Belusca-Maito:2014dpa}. The only remaining divergences are the logarithmic ones, which are constrained using electroweak precision measurements.

	\par We are thus left with the following Higgs effective Lagrangian, written, after EWSB and in unitary gauge, as an expansion in powers of the Higgs boson $h$:
	\begin{equation}
		\mathcal{L}_{eff} = \mathcal{L}_0 + \mathcal{L}_h + \cdots
	\end{equation}
	where $\mathcal{L}_0$ is the Higgs-independent part and $\mathcal{L}_h$ is the phenomenological Higgs Lagrangian. It should be noted that the dimension-6 effective Lagrangian gives also contributions to $\mathcal{L}_0$~\cite{Belusca-Maito:2014dpa}. We stop at the first power in $h$ because current LHC constraints for couplings with more than one Higgs boson are weak. For $\mathcal{L}_h$ we obtain:
	\begin{equation}\begin{split}
		\mathcal{L}_h = \frac{h}{v} \left[\vphantom{\sum}\right.
		& 2 c_V m_W^2 W_{\mu}^\dagger W^{\mu} + c_V m_Z^2 Z_{\mu} Z^{\mu} \\
		& - \sum_{f=u,d,l} m_f \overline{f} \left(c_f + \imath\gamma_5\,\widetilde{c_f}\right) f \\
		& - \frac{1}{2} c_{WW} W_{\mu\nu}^\dagger W^{\mu\nu} - \frac{1}{2} \widetilde{c}_{WW} W_{\mu\nu}^\dagger \widetilde{W}^{\mu\nu} \\
		& - \frac{1}{4} c_{ZZ} Z_{\mu\nu} Z^{\mu\nu} - \frac{1}{4} \widetilde{c}_{ZZ} Z_{\mu\nu} \widetilde{Z}^{\mu\nu} \\
		& - \frac{1}{4} c_{\gamma\gamma} \gamma_{\mu\nu} \gamma^{\mu\nu} - \frac{1}{4} \widetilde{c}_{\gamma\gamma} \gamma_{\mu\nu} \widetilde{\gamma}^{\mu\nu} \\
		& - \frac{1}{2} c_{Z\gamma} \gamma_{\mu\nu} Z^{\mu\nu} - \frac{1}{2} \widetilde{c}_{Z\gamma} \gamma_{\mu\nu} \widetilde{Z}^{\mu\nu} \\
		& + \frac{1}{4} c_{gg} G^a_{\mu\nu} G^{a{\mu\nu}} + \frac{1}{4} \widetilde{c}_{gg} G^a_{\mu\nu} \widetilde{G}^{a{\mu\nu}}
		\left.\vphantom{\sum}\right]
	\end{split}\end{equation}
	where we used the field-strength tensors $V_{\mu\nu}$ and their duals $\widetilde{V}_{\mu\nu} \equiv \frac{1}{2} \epsilon_{\mu\nu\rho\sigma} V^{\rho\sigma}$. This Lagrangian only depends on 7 independent parameters in the CP-even sector:
	\begin{equation}
		c_V, \quad c_u, \quad c_d, \quad c_l, \quad c_{\gamma\gamma}, \quad c_{Z\gamma}, \quad c_{gg}
	\end{equation}
	and 6 independent parameters in the CP-odd sector:
	\begin{equation}
		\widetilde{c}_u, \quad \widetilde{c}_d, \quad \widetilde{c}_l, \quad \widetilde{c}_{\gamma\gamma}, \quad \widetilde{c}_{Z\gamma}, \quad \widetilde{c}_{gg}
	\end{equation}
	and the (tree-level) SM Higgs Lagrangian is retrieved when $c_V = c_{f=u,d,l} = 1$, $c_{gg} = c_{\gamma\gamma} = c_{Z\gamma} = 0$ and all the $\widetilde{c}_i = 0$. They can be however generated at loop-level in the SM: for instance the $c_{\gamma\gamma}$ coefficient for the $h \gamma \gamma$ coupling is generated via a fermion or boson loop. In BSM models such as two-Higgs-doublet models (2HDM), new contributions can arise, for example a loop of charged Higgs bosons. CP-odd couplings, which cannot arise from the SM, can also be generated if a pseudo-scalar Higgs boson is present.

\section{Constraints on Higgs couplings}
	\par We use the Higgs signal strengths that are usually provided by the ATLAS and CMS experiments in various Higgs channels, defined as: $\hat{\mu}^{YH}_{XX} = \frac{\sigma_{YH}}{\sigma_{YH}^{SM}} \frac{{\rm Br}(h \to XX)}{{\rm Br}(h \to XX)_{SM}}$, as well as the 2-dimensional likelihood functions defined in the $\hat\mu_{ggH+ttH}$--$\hat\mu_{\text{VBF+VH}}$ plane for different Higgs channels. Those 2D likelihoods are very useful because they encode the non-trivial correlations between the rates measured for the $ggH$/$ttH$ or VBF/VH production modes. Also constraints from EW precision data from LEP, SLC and Tevatron are used (see Section~4 and Table~3 in~\cite{Belusca-Maito:2014dpa} for more details). Within the effective Higgs theory we compute the corresponding relative branching fractions and production cross-sections and we perform a global fit with respect to the experimental signal strengths.

	\par
\begin{figure*}
		\centering
		\subcaptionbox{}{\includegraphics[width=0.3\textwidth]{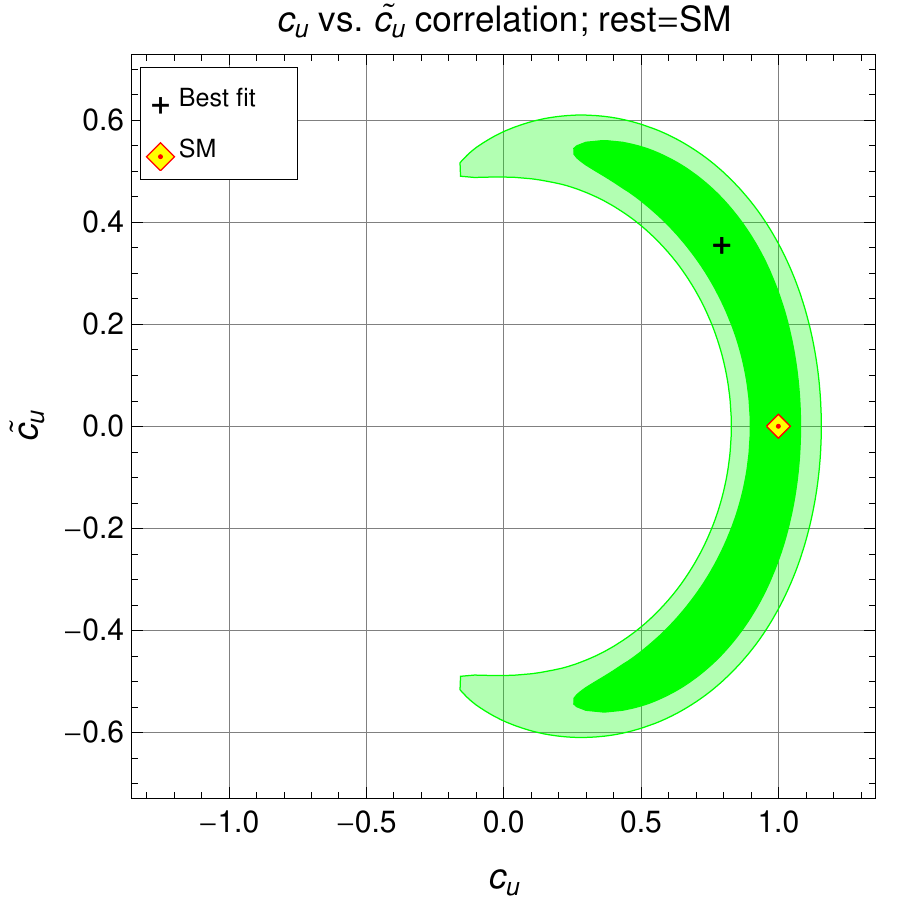}} \hspace{0.3cm}
		\subcaptionbox{}{\includegraphics[width=0.3\textwidth]{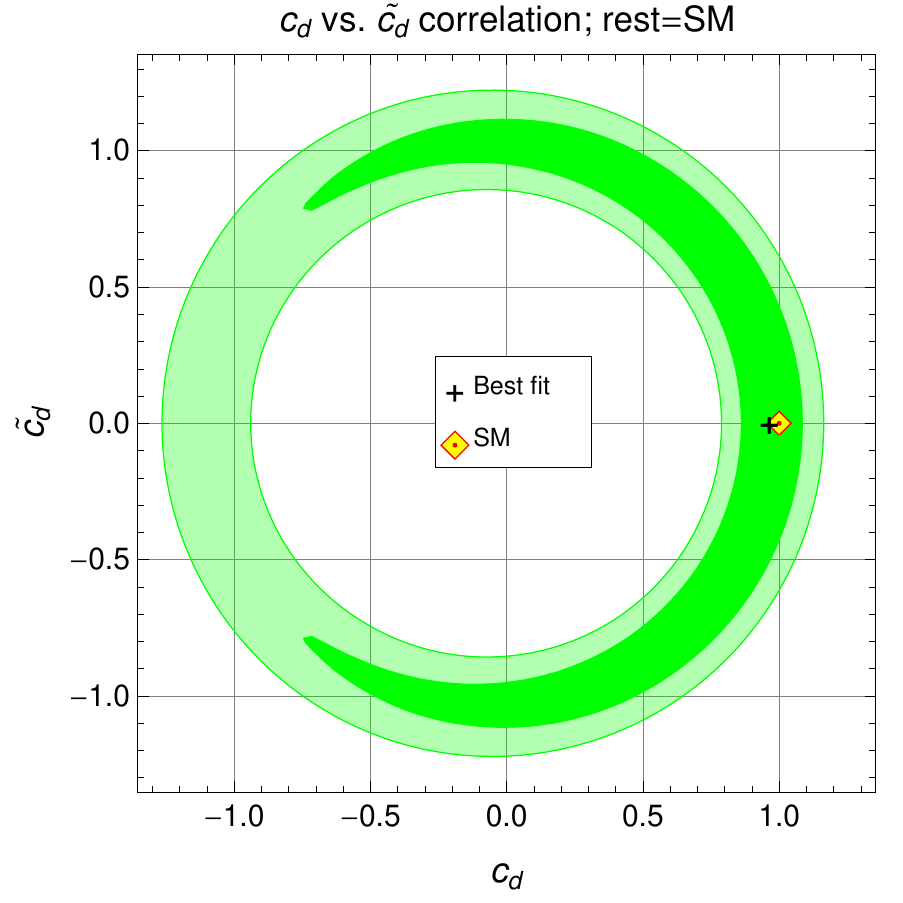}} \hspace{0.3cm}
		\subcaptionbox{}{\includegraphics[width=0.3\textwidth]{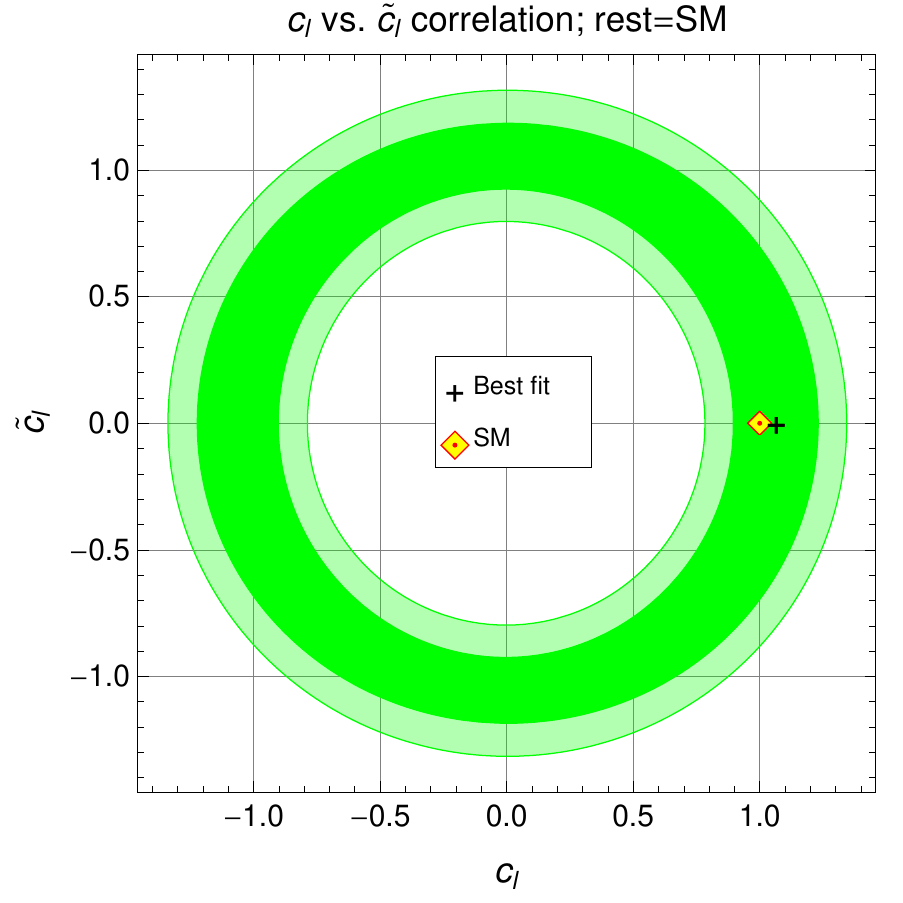}}
		\caption{Fits in the $c_u$--$\widetilde{c}_u$ (left), $c_d$--$\widetilde{c}_d$ (center) and $c_l$--$\widetilde{c}_l$ (right) planes with the other couplings fixed to their SM values. Dark green: $68\%$~CL; light green: $95\%$~CL. The displayed best-fit points are found for a fit around the SM values. For lepton couplings we obtain a complete degeneracy, partially lifted for up-type quarks.}
		\label{fig:ci_cit}
	\end{figure*}

 We obtain the following central values and 68\%~CL intervals for the effective couplings. For the CP-even couplings:
	\begin{eqnarray*}
		& c_V = 1.04 \pm 0.03, \quad c_l = 1.09^{+0.13}_{-0.11}, \nonumber \\
		& c_u = 1.31^{+0.10}_{-0.34}, \quad c_d = 0.92^{+0.22}_{-0.13}, \nonumber \\
		& c_{gg} = -0.0016^{+0.0021}_{-0.0022}, \quad c_{\gamma\gamma} = 0.0009^{+0.0008}_{-0.0010}, \nonumber \\
		& c_{Z\gamma} = -0.0006^{+0.0183}_{-0.0240},
	\end{eqnarray*}
	and for the CP-odd ones:
	\begin{eqnarray*}
		& \widetilde{c}_u = \pm (0.87^{+0.33}_{-2.08}), \quad \widetilde{c}_d = -0.0035^{+0.4608}_{-0.4581}, \nonumber \\
		& \widetilde{c}_l = \pm (0.37^{+0.25}_{-0.99}), \quad \widetilde{c}_{gg} = 0.0004^{+0.0038}_{-0.0040}, \nonumber \\
		& \widetilde{c}_{\gamma\gamma} = \pm (0.0033^{+0.0017}_{-0.0028}), \quad \widetilde{c}_{Z\gamma} = 0.0075^{+0.0200}_{-0.0345}.
	\end{eqnarray*}
	A $\chi_{\rm SM}^2 - \chi_{\rm min}^2 = 5.3$ is obtained, meaning the SM gives a perfect fit to the Higgs and EW precision data. We notice the current data already place meaningful limits on the seven CP-even parameters. The least stringent constraint is the one on $c_{Z\gamma}$ that reflects the current weak experimental limits on the $h \to Z \gamma$ decay rate; however this can be improved by using differential cross-section measurements in the so-called "Golden Channel" $h \to 4\ell$~\cite{Chen:2013ejz,Chen:2014pia,Chen:2014gka}. Concerning the CP-odd couplings, if the Higgs-gauge couplings $\widetilde{c}_{gg}$, $\widetilde{c}_{\gamma\gamma}$ and $\widetilde{c}_{Z\gamma}$ are correctly constrained and are all compatible with zero, the up-type and leptonic couplings are not very well constrained and have a sign degeneracy: the Higgs rate measurements indeed constrain only the sum of the squares of the CP-even and CP-odd couplings (or a combination thereof) and not their signs, as shown in Fig.~\ref{fig:ci_cit}. A way to improve the precision on those couplings would be to perform other studies such as looking at the differential cross-section measurements or to use electric dipole moments of the electron or the neutron~\cite{Brod:2013cka} used together with 14~TeV LHC data at 3000~fb$^{-1}$.

\section{Summary}
	\par We reanalyzed constraints on Higgs couplings to matter using an effective theory framework using effective dimension-6 operators, from which the phenomenological Higgs Lagrangian was rederived. LHC Higgs rates data and electroweak precision measurements from LEP, SLC and Tevatron allowed us to obtain strong constraints on CP-even parameters that are found to be compatible with SM within 68\%~CL. The CP-odd parameters are less constrained and some of them show a sign degeneracy: this fact is understandable because Higgs rates constrain a combination of the square of these couplings; nevertheless the constraints are perfectly compatible with SM values. It is expected that using electric dipole moments or tensor structures together with new data from the next run of the LHC should greatly improve those constraints.

	\par Higgs-mediated flavour violation operators were explicitely removed from this study; however they constitute a large number of possible dimension-6 operators, basically the Yukawa-like operators as well as the 2-fermion-Higgs vertex and dipole operators. They are currently analyzed~\cite{Belusca-Maito:InPreparation}. The aim would be to constrain their couplings using indirect and direct limits and get upper bounds on possible new Higgs exotic processes that may be seen at the next LHC run.

\renewcommand\refname{References}


\input{hbelusca.bbl}

%% file: session/hbelusca/hbelusca.bbl
\providecommand{\href}[2]{#2}

%% file: hbelusca_JJC2014_arxiv.bbl
\begingroup\begin{thebibliography}{10}

\bibitem{Aad:2012tfa}
{\bfseries ATLAS Collaboration}, G.~Aad {\em et~al.}, ``{Observation of a new
  particle in the search for the Standard Model Higgs boson with the ATLAS
  detector at the LHC}'',
  \href{http://dx.doi.org/10.1016/j.physletb.2012.08.020}{{\em Phys.Lett.}
  {\bfseries B716} (2012) 1--29},
\href{http://arxiv.org/abs/1207.7214}{{\ttfamily arXiv:1207.7214 [hep-ex]}}.

\bibitem{Chatrchyan:2012ufa}
{\bfseries CMS Collaboration}, S.~Chatrchyan {\em et~al.}, ``{Observation of a
  new boson at a mass of 125 GeV with the CMS experiment at the LHC}'',
  \href{http://dx.doi.org/10.1016/j.physletb.2012.08.021}{{\em Phys.Lett.}
  {\bfseries B716} (2012) 30--61},
\href{http://arxiv.org/abs/1207.7235}{{\ttfamily arXiv:1207.7235 [hep-ex]}}.

\bibitem{Englert:1964et}
F.~Englert and R.~Brout, ``{Broken Symmetry and the Mass of Gauge Vector
  Mesons}'',
\href{http://dx.doi.org/10.1103/PhysRevLett.13.321}{{\em Phys.Rev.Lett.}
  {\bfseries 13} (1964) 321--323}.

\bibitem{Higgs:1964pj}
P.~W. Higgs, ``{Broken Symmetries and the Masses of Gauge Bosons}'',
\href{http://dx.doi.org/10.1103/PhysRevLett.13.508}{{\em Phys.Rev.Lett.}
  {\bfseries 13} (1964) 508--509}.

\bibitem{Guralnik:1964eu}
G.~Guralnik, C.~Hagen, and T.~Kibble, ``{Global Conservation Laws and Massless
  Particles}'',
\href{http://dx.doi.org/10.1103/PhysRevLett.13.585}{{\em Phys.Rev.Lett.}
  {\bfseries 13} (1964) 585--587}.

\bibitem{Grzadkowski:2010es}
B.~Grzadkowski, M.~Iskrzynski, M.~Misiak, and J.~Rosiek, ``{Dimension-Six Terms
  in the Standard Model Lagrangian}'',
  \href{http://dx.doi.org/10.1007/JHEP10(2010)085}{{\em JHEP} {\bfseries 1010}
  (2010) 085},
\href{http://arxiv.org/abs/1008.4884}{{\ttfamily arXiv:1008.4884 [hep-ph]}}.

\bibitem{Belusca-Maito:InPreparation}
H.~Belusca-Maito and A.~Falkowski. \emph{In preparation}.

\bibitem{Belusca-Maito:2014dpa}
H.~Belusca-Maito, ``{Effective Higgs Lagrangian and Constraints on Higgs
  Couplings}'',
\href{http://arxiv.org/abs/1404.5343}{{\ttfamily arXiv:1404.5343 [hep-ph]}}.

\bibitem{Peskin:1991sw}
M.~E. Peskin and T.~Takeuchi, ``{Estimation of oblique electroweak
  corrections}'',
\href{http://dx.doi.org/10.1103/PhysRevD.46.381}{{\em Phys.Rev.} {\bfseries
  D46} (1992) 381--409}.

\bibitem{Baak:2012kk}
{\bfseries The GFitter Group}, M.~Baak, M.~Goebel, J.~Haller, A.~Hoecker,
  D.~Kennedy, R.~Kogler, K.~Moenig, M.~Schott, and J.~Stelzer, ``{The
  Electroweak Fit of the Standard Model after the Discovery of a New Boson at
  the LHC}'', \href{http://dx.doi.org/10.1140/epjc/s10052-012-2205-9}{{\em
  Eur.Phys.J.} {\bfseries C72} (2012) 2205},
\href{http://arxiv.org/abs/1209.2716}{{\ttfamily arXiv:1209.2716 [hep-ph]}}.

\bibitem{Chen:2013ejz}
Y.~Chen and R.~Vega-Morales, ``{Extracting Effective Higgs Couplings in the
  Golden Channel}'',
\href{http://arxiv.org/abs/1310.2893}{{\ttfamily arXiv:1310.2893 [hep-ph]}}.

\bibitem{Chen:2014pia}
Y.~Chen, E.~Di~Marco, J.~Lykken, M.~Spiropulu, R.~Vega-Morales, {\em et~al.},
  ``{8D Likelihood Effective Higgs Couplings Extraction Framework in the Golden
  Channel}'',
\href{http://arxiv.org/abs/1401.2077}{{\ttfamily arXiv:1401.2077 [hep-ex]}}.

\bibitem{Chen:2014gka}
Y.~Chen, R.~Harnik, and R.~Vega-Morales, ``{Probing the Higgs Couplings to
  Photons in $h\rightarrow 4\ell$ at the LHC}'',
\href{http://arxiv.org/abs/1404.1336}{{\ttfamily arXiv:1404.1336 [hep-ph]}}.

\bibitem{Brod:2013cka}
J.~Brod, U.~Haisch, and J.~Zupan, ``{Constraints on CP-violating Higgs
  couplings to the third generation}'',
  \href{http://dx.doi.org/10.1007/JHEP11(2013)180}{{\em JHEP} {\bfseries 1311}
  (2013) 180},
\href{http://arxiv.org/abs/1310.1385}{{\ttfamily arXiv:1310.1385 [hep-ph]}}.

\end{thebibliography}\endgroup
